\documentclass[conference]{IEEEtran}

\IEEEoverridecommandlockouts

\usepackage{cite}
\usepackage{amsmath,amssymb,amsfonts}
\usepackage{graphicx}
\usepackage{textcomp}
\usepackage{xcolor}
\usepackage{bm}
\usepackage{algorithm}
\usepackage{subfigure}

\def\BibTeX{{\rm B\kern-.05em{\sc i\kern-.025em b}\kern-.08em
    T\kern-.1667em\lower.7ex\hbox{E}\kern-.125emX}}
\begin{document}
\title{Uncertainty in Position Estimation Using Machine Learning}

\author{\IEEEauthorblockN{Yuxin Zhao}
\IEEEauthorblockA{yuxin.zhao@ericcson.com\\
\textit{Ericsson AB}\\
Linköping, Sweden \\
}
\and
\IEEEauthorblockN{Deep Shrestha}
\IEEEauthorblockA{deep.shrestha@ericsson.com\\
\textit{Ericsson AB}\\
Linköping, Sweden \\
}
}

\maketitle

\begin{abstract}
UE localization has proven its implications on multitude of use cases ranging from emergency call localization to new and emerging use cases in industrial IoT. To support plethora of use cases Radio Access Technology (RAT)-based positioning has been supported by 3GPP since Release 9 of its specifications that featured basic positioning methods based on Cell Identity (CID) and Enhanced-CID (E-CID). Since then, multiple positioning techniques and solutions are proposed and integrated in to the 3GPP specifications. When it comes to evaluating performance of the positioning techniques, achievable accuracy (2-Dimensional or 3-Dimensional) has, so far, been the primary metric. With an advent of Release 16 New Radio (NR) positioning, it is possible to configure Positioning Reference Signal (PRS) with wide bandwidth that naturally helps improving the positioning accuracy. However, the improvement is evident when the conditions are ideal for positioning. In practice where the conditions are non-ideal and the positioning accuracy is severely impacted, estimating the uncertainty in position estimation becomes important and can provide significant insight on how reliable a position estimation is.

In order to determine the uncertainty in position estimation we resort to Machine Learning (ML) techniques that offer ways to determine the uncertainty/reliability of the predictions for a trained model. Hence, in this work we propose to combine ML methods such as Gaussian Process (GP) and Random Forest (RF) with RAT-based positioning measurements to predict the location of a UE and in the meantime also assess the uncertainty of the estimated position. The results show that both GP and RF not only achieve satisfactory positioning accuracy but also give a reliable uncertainty assessment of the predicted position of the UE.
\end{abstract}


\section{Introduction}
Since its inception Radio Access Technology (RAT) has mainly been developed and deployed to provide seamless communication services to end consumers \cite{8667173}. The communication services since then have evolved and range today from provisioning a voice call setup between two end users to providing required connectivity to mobile broadband users, including but not limited to Device-to-Device communication, and vehicular use cases using wireless radio signals \cite{7970319, 9217500}. As new use cases are emerging, the RAT-based technology is compelling in itself to provide services beyond communication \cite{4127518}. An example of such a service is positioning. Positioning has been a key area of investigation since Release 9 of Third Generation Partnership Project (3GPP) specification and is evolving towards providing positioning as a service to localize both indoor and outdoor User Equipments (UEs) and can overcome drawbacks that other positioning solutions hold \cite{8377447}. For instance, satellite-based positioning systems can only localize outdoor UEs and the achievable accuracy typically is limited to visibility of multiple satellite stations at the UE location \cite{9053157}. Ultra Wide Band (UWB) positioning solutions can achieve decent positioning accuracy within a local area where UWB sensors are deployed and therefore is not a solution that can offer wide area positioning as a service \cite{8964963}.

RAT supports a multitude of positioning techniques namely: Cell Identity (CID) positioning, Enhanced CID (E-CID) positioning, Observed Time Difference of Arrival (OTDoA) based positioning, Round Trip Time (RTT) based positioning etc \cite{8877160}. In CID-based positioning, the serving cell location is estimated to be the UE location. Since cell radius may vary from a few meters (20 meters in an indoor deployment) to a few kilometers (7 km in rural urban macro deployment) depending on the deployment type, the CID positioning accuracy therefore varies from meters to kilometers. Achievable accuracy of such a positioning technique can be enhanced by using complementary information such as Timing Advance (TA) \cite{7217158}. The positioning method that combines the complementary information with the serving cell location is known as ECID. Since TA values are calculated with lower resolution the positioning accuracy of ECID method is higher than CID positioning but is limited to estimating the UE location as a circle whose radius is equal to the range given by the TA value associated to the UE to be positioned. With the rising interest on RAT-based positioning from many verticals that require higher accuracy than the ones that are achievable by the above-mentioned methods both Long Term Evolution (LTE) and New Radio (NR) support positioning methods that exploit OTDoA and RTT measurements for UE localization. These methods achieve higher positioning accuracy in comparison to the CID and E-CID methods but are limited to geometry of deployment, time synchronization between the UE and the Base Station (BS) and/or time synchronization among the BSs. Geometry of deployment refers to the location of BSs transmitting Positioning Reference Signal (PRS) during a positioning occasion and can dilute the precision of achievable positioning accuracy of OTDoA positioning method \cite{8970252}. Moreover, the achievable positioning accuracy of OTDoA method also heavily depends on the timing synchronization among the BSs transmitting PRSs. An un-synched transmission of PRSs will affect the Reference Signal Time Difference (RSTD) measurement (performed either by the UE or the BS depending on which node is transmitting the PRS) that eventually results into erroneous UE position estimation. The error on UE position estimation due to un-synched network can be reduced significantly by using the RTT measurements instead of RSTD measurements. However, single cell RTT requires complementary angular measurement to estimate UE location and the multi-cell RTT method relies on geometry of the deployment for high accuracy UE localization \cite{2021arXiv210203361D}. Besides the timing measurement-based positioning methods are also susceptible to Time of Arrival (ToA) measurement error, due to Non-Line-of-Sight (NLoS) condition of the radio link, that has implication on the achievable positioning accuracy \cite{8417725}. Overcoming all these issues to enhance the positioning accuracy of the RAT-based solution is not trivial. Therefore, it is important to evaluate the uncertainty of the estimated UE position. Supplementing such information will enhance the adoptability of RAT-based positioning methods to support use cases that demand stringent positioning accuracy.

Assessing the uncertainty/reliability of the position estimation in the RAT-based positioning framework is feasible with the help of Machine Learning (ML) methods. In machine learning, there exist different ways to determine the uncertainty of the predictions for a trained model. In this work, we propose two different ML methods to assess the positioning uncertainty associated with the estimation. One is to combine Gaussian Process (GP) regression with RAT-based positioning method. The other one is to use Random Forest (RF) for both position estimation and uncertainty assessment. 

To give an overview of the problem and the proposed solution Downlink (DL)-OTDoA is taken as an example positioning method to perform the subsequent analysis. For concise elaboration, the remainder of this paper is divided into four sections. Section \ref{sec:otdoa} will give a brief introduction to DL-OTDoA method. Section \ref{sec:ml_uncertainty} will elaborate on ML method for uncertainty assessment. Section \ref{sec:results} will showcase results from numerical validation. And finally Section \ref{sec:conclusions} will present concluding remarks. 

\section{DL-OTDoA Positioning}\label{sec:otdoa}
\begin{figure}[h]
\includegraphics[width = \linewidth]{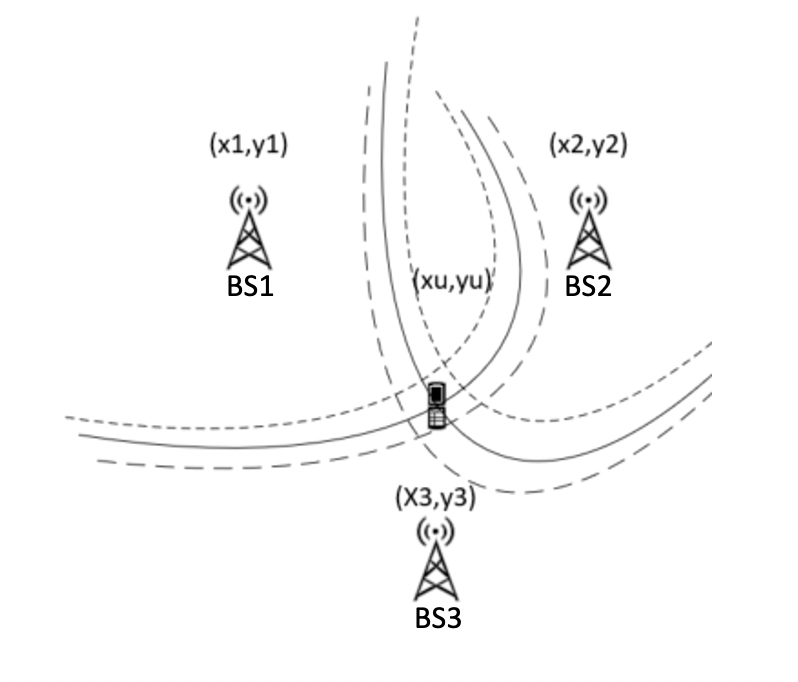}
\caption{OTDOA based positioning.}
\label{multilateration}
\end{figure}

In DL-OTDoA based positioning, a UE performs RSTD measurements which is the time difference of arrival of the PRSs transmitted from a reference BS and multiple neighbouring BSs. For UE localization typically two such measurements are required. The RSTD measurements are then used to perform multilateration to resolve UE location by drawing hyperbolas as shown in Fig.\ref{multilateration}. The intersecting point of such hyperbolas is the estimated location of the UE. 

Assuming a 3 BS deployment as shown in Fig.\ref{multilateration}, where BS1 is located at $(x_1,y_1)$, BS2 is located at $(x_2,y_2)$, BS3 is located at $(x_3,y_3)$, the RSTD measurement (under ideal condition), which is the difference of ToA from BS2 (reference BS) and BS1 at UE location $(x_u,y_u)$ can be expressed as shown in \eqref{eqn:1}.
\begin{equation}\label{eqn:1}
{\textrm{RSTD}_{2,1}} = {\tau_2 - \tau_1},
\end{equation}
where
\begin{equation*}
{\tau_2} = 
\frac{\sqrt{(x_u-x_2)^2  + (y_u -y_2) ^2}}{c}
\end{equation*}
and
\begin{equation*}
{\tau_1} =
\frac{\sqrt{(x_u-x_1)^2  + (y_u -y_1) ^2}}{c}
\end{equation*}
are ToA measurements from BS2 and BS1 and $c$ denotes the speed of the light.
Likewise, RSTD measurement taking BS2 as a reference BS and BS3 as a neighbouring BS can be expressed as shown in \eqref{eqn:2}
\begin{equation}\label{eqn:2}
{\textrm{RSTD}_{2,3}} = {\tau_2 - \tau_3},
\end{equation}
where
\begin{equation*}
{\tau_3} = 
\frac{\sqrt{(x_u-x_3)^2  + (y_u -y_3) ^2}}{c}.
\end{equation*}
Under ideal condition when the BSs transmitting PRS are perfectly time synched, the UE is in Line-of-Sight (LoS) to all BSs and the RSTD measurements are perfect, the hyperbolas for multilateration are the solid line as shown in Fig.\ref{multilateration}. These solid lines intersect at the UE location allowing a precise estimation of UE position. However, when the PRS transmitting BSs are un-synched and the RSTD measurements are affected by erroneous ToA estimate due to NLoS condition between the UE and the BSs, the hyperbolas for multilateration do not intersect at UE location leading to a high positioning error \cite{otdoa_sven}. In such a situation, it is of utmost importance that the uncertainty in estimated position is evaluated and is used as an additional information to guarantee reliability of positioning estimate.   

\section{Machine Learning for Uncertainty Assessment}\label{sec:ml_uncertainty}
Assessing the uncertainty/reliability of the position estimation in the RAT-based positioning framework is feasible with the help of ML methods. In ML, there exist different ways to determine the uncertainty/reliability of the predictions for a trained model\cite{BOSNIC2008504}. Hence, we propose to combine ML methods with RAT-based positioning measurements to predict the location of a UE and in the meantime, assess the uncertainty of the position estimate. In the subsequent sections, two methods will be introduced.  

\subsection{Gaussian Process for Uncertainty Assessment}
Gaussian Process (GP) in known as a Bayesian non-parametric ML approach. The gist of GP models is to impose a Gaussian prior on the function/system $f(x)$ and then compute the posterior distribution over the function given the observed data. GP models have been used in a plethora of applications due to their outstanding performance in function approximation with a self-contained uncertainty bound. One advantage of using GP is that there is no need to have prior assumptions about the function itself. And the function is fully determined by the training data set. To be more specific, a GP is a generalization of the Gaussian probability distribution. In a GP, every point in some continuous input space is associated with a normally distributed random variable. Moreover, every finite collection of those random variables has a multivariate normal distribution. The distribution of a GP is the joint distribution of all those (infinitely many) random variables.

Consider ToA used in this paper. The distance between UE and the BS can be estimated using the ToA measurements. However, the direct estimation, which takes the ToA divided by the speed of the light, is usually not accurate in practical situations where the ToA measurements are noisy and corrupted due to misdetection of the first arrival path in a multi path propagation environment. Moreover, if the UE is in NLoS condition with the PRS transmitting BS the first arrival path usually induces excess timing delay such that the ToA estimates do not correspond to time of flight of signal equivalent to the geometric distance between the UE and the BS. Hence, we propose to model the distance from a UE to the $i$-th BS as a function of ToA measurements. If $\tau_i$ denotes the ToA measurement to the $i$-th BS, then a generic model can be given as
\begin{equation}
d_i=f_i (\tau_i )+n_i,
\end{equation}
where $n_i$ is assumed to be zero mean Gaussian noise with variance $\sigma_i^2$, and the function $f_i(\tau_i)$ is modeled by a GP that can be formulated as
\begin{equation}\label{GP_eqn}
f_i(\tau_i)\sim \mathcal{GP}(m_i(\tau_i),k_i(\tau_i,\tau_i')). 
\end{equation}
In \eqref{GP_eqn}, the variable $m_i(\tau_i)$ is the mean function, and $k_i(\tau_i,\tau_i')$ is the covariance/kernel function which indicates the correlation between two ToA measurements $\tau_i$ and $\tau_i'$. If we rewrite the distances from the UE to $N$ BSs in vector form, we have
\begin{equation}
\bm{d} = \bm{f}(\bm{\tau})+\bm{n},
\end{equation} 
where $\bm{d}=[d_1, d_2,\ldots,d_N]^T$, $\bm{f}(\bm{\tau})=[f_1(\tau_1),\ldots,f_N(\tau_N)]^T$, and $\bm{n}=[n_1, \ldots,n_N]^T$. The notation $[.]^T$ denotes transpose of the vector $[.]$.

In the training phase, adequate ToA measurements are collected together with the positions where the measurements are done. Then, GPs are trained for each BS. Let us take the $i$-th BS as one example and further ignore the index for simplicity. With the training data, denote the distances from different locations to the BS as $\bar{\bm{d}} = [\bar{d}_0, \bar{d}_1, \ldots, \bar{d}_{M-1}]^T$, where $M$ is the number of collected measurements. Correspondingly, the training ToA are denoted as $\bar{\bm{\tau}}$. We can further denote the training data set as $\mathcal{D}\triangleq\left\{\bar{\bm{\tau}}, \bar{\bm{d}}\right\}$. In GP, the joint distribution of the distances is still Gaussian \cite{RW06}. Hence, we have
\begin{equation}
\label{eq:likelihood}
	p(\bar{\bm{d}}|\bar{\bm{\tau}})\sim \mathcal{N}(\bm{m}(\bar{\bm{d}}),K(\bar{\bm{\tau}},\bar{\bm{\tau}})+\sigma^2I_M),
\end{equation}
where $\bm{m}(\bar{\bm{d}})=[m(\bar{d}_0),\ldots,m(\bar{d}_{M-1})]^T$, $I_M$ is identity matrix of size $M$, and $K(\bar{\bm{\tau}},\bar{\bm{\tau}})$ is the kernel matrix, which is given by:
\begin{equation}
K(\bar{\bm{\tau}},\bar{\bm{\tau}}) =  
\left[\begin{array}{ccc}
k(\bar{\tau}_0,\bar{\tau}_0) & \cdots & k(\bar{\tau}_0,\bar{\tau}_{M-1}) \\
\vdots & \ddots & \vdots \\
k(\bar{\tau}_{M-1},\bar{\tau}_0) & \cdots & k(\bar{\tau}_{M-1},\bar{\tau}_{M-1}) 
\end{array}\right].
\end{equation}
Parameters such as $\sigma$, and hyperparameters in the kernel function can be learned during the training phase by maximizing the log-likelihood function $\log p(\bar{\bm{d}}|\bar{\bm{\tau}})$ according to \cite{RW06}.

In the prediction/positioning phase, we make use of the trained GP models to predict the distances from the UE to BSs based on the ToA estimations. Assume the new ToA estimation is denoted as $\tau^*$, the predicted distance to a specific BS $d^*$ is estimated as
\begin{equation}
p(d^*|\tau^*, \mathcal{D}) \sim \mathcal{N}(\hat{\mu}(d^*), \hat{k}(d^*)),\label{eq:predDist}
\end{equation} 
where by using a zero mean function $m(\tau)=0$, 
\begin{align}
\hat{\mu}(\tau^*) &= k(\tau^*,\bar{\bm{\tau}})^T (K(\bar{\bm{\tau}},\bar{\bm{\tau}})+\sigma^2 I_M)^{-1} \bar{\bm{d}},\nonumber \\
 \hat{k}(\tau^*) &= k(\tau^*,\tau^*)+\sigma^2\nonumber \\
 &-k(\tau^*,\bar{\bm{\tau}})^T (K(\bar{\bm{\tau}},\bar{\bm{\tau}})+\sigma^2 I_M)^{-1}k(\tau^*,\bar{\bm{\tau}}),\nonumber
\end{align}
where $k(\tau^*,\bar{\bm{\tau}})=[k(\tau^*,\bar{\tau}_0), \ldots, k(\tau^*,\bar{\tau}_{M-1})]^T$. The detailed derivations can be found in \cite{RW06}. From \eqref{eq:predDist}, we see that the predicted distance from UE to each BS is Gaussian distributed with a mean and variance value, which can be denoted as:
\begin{equation}
\hat{d}_i\sim\mathcal{N}(\hat{\mu}(\tau_i),\hat{k}(\tau_i)), i = 1, \ldots, N,
\end{equation}
where $\tau_i $ is the ToA measurement of the $i$-th BS. 

With above derivations, one algorithm is designed to estimate the position of UE and assess the uncertainty of the position estimates based on the predicted distances, as provided in Algorithm \ref{alg:Algorithm2}. Here the algorithm works with the assumption that GP models have been trained off-line, and the ToA measurements obtained from the radio network are denoted as $\bm{\tau}=[\tau_1, \ldots,\tau_N]^T$ without considering any uncertainty in the ToA estimations.  

%

%

\begin{algorithm}[tb]
	\caption{Sampling-based Uncertainty Assessment}
	\label{alg:Algorithm2}
	\begin{enumerate}
		\item The distances from the UE to all BSs are computed as: $d_i = \hat{\mu}(\tau_i)$, for $i=1, \ldots,N$.
		\item Compute the position estimation as
		\begin{equation}
			\hat{\bm{p}} =  \textrm{OTDoA}(\hat{\mu}(\bm{\tau})),
		\end{equation}
		where $\hat{\mu}(\bm{\tau})=[\hat{\mu}(\tau_1),\ldots,\hat{\mu}(\tau_N)]^T$, and \textrm{OTDoA} denotes the positioning method introduced in Section \ref{sec:otdoa}.
		\item For $n_s = 1, \ldots,N_s$, do
		\begin{itemize}
			\item Take samples from the Gaussian distribution $\mathcal{N}(\hat{\mu}(\tau_i),\hat{k}(\tau_i))$, for $i=1, \ldots,N$. Denote the sampled distances as ${d_{n_s}}_i$. 
			\item Use ${d_{n_s}}_i$, $i=1, \ldots,N$ to compute the position estimate $\hat{\bm{p}}_{n_s}$ according to the positioning method in Section \ref{sec:otdoa}. 
		\end{itemize} 
		\item Compute the uncertainty as:
		\begin{equation}
		\bm{v}(\hat{\bm{p}}) = \frac{1}{N_s-1}\sum_{n_s=1}^{N_s}(\hat{\bm{p}}_{n_s}-\hat{\bm{p}})^2.
		\end{equation}
	\end{enumerate}
\end{algorithm}

\subsection{Random Forest for Uncertainty Assessment}
RF is an ensemble of decision trees. RF applies the general technique of bootstrap aggregating, or bagging, to decision trees (i.e., each decision tree selects a random sample with replacement of the training data set and fits the decision tree with these selected samples). The splitting of each decision tree is learned during the training phase.

\begin{figure}[t]
	\centering
	\includegraphics[width=\linewidth]{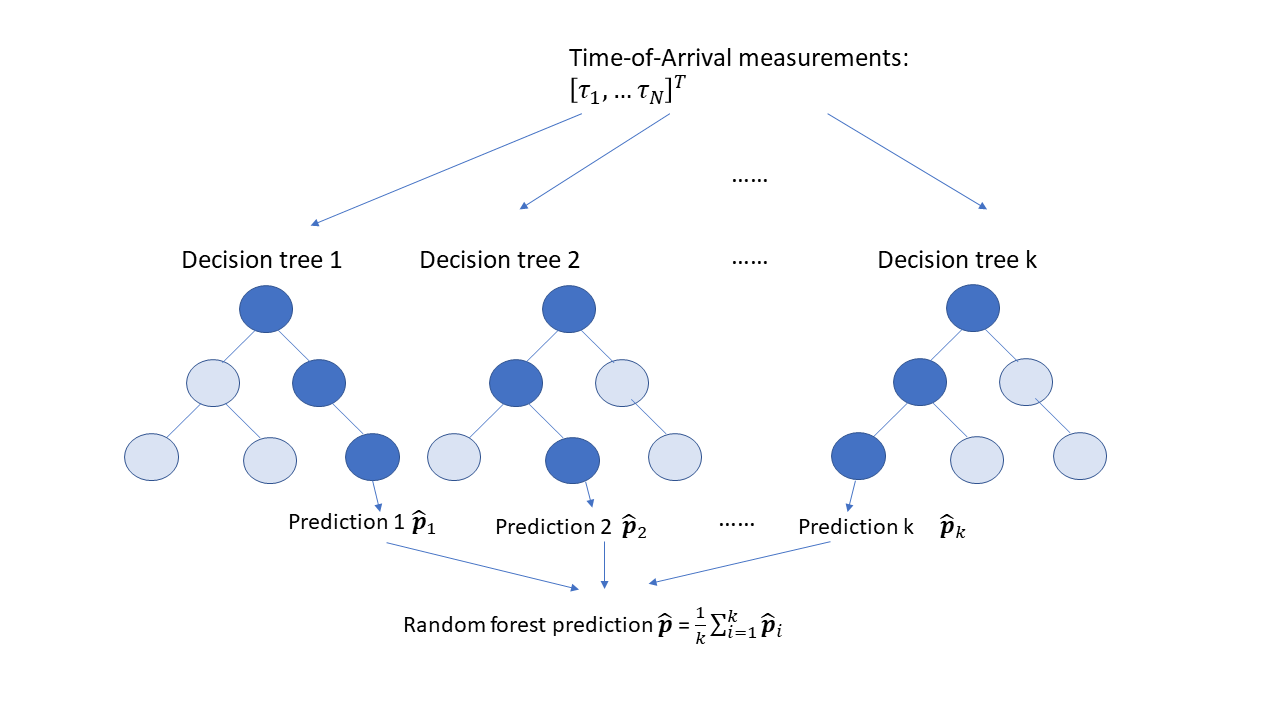}
	\caption{Illustration of Random Forest for position estimation.}
	\label{fig:fig1}
\end{figure}

To perform position estimation and uncertainty assessment, RF is applied as the ML model to find the relationship between the position of the UE and corresponding radio measurements. Again we take ToA as one example. The position can be learned from the the ToA measurements from different BSs. Correspondingly, we have
\begin{equation}
\bm{p} = f(\bm{\tau}),
\end{equation}
where $\bm{p}$ denotes the output which includes the 2-Dimensional position, $\bm{\tau}=[\tau_1,\ldots, \tau_N]^T$ is the input feature which includes ToA measurements from all BSs, and $f$ denotes the ML model, which is learned using RF in this particular case. The structure of RF for position estimation can be illustrated in Fig.~\ref{fig:fig1}.

To train the RF model, a set of training data is required. The data can be either collected by field trials for positioning purposes or the network can explore potential measurements that are not intentionally designed for positioning. In the latter case, the network would identify the positions at which those measurements are collected. In the on-line position prediction phase, given the trained RF model and new ToA measurements $\bm{\tau}^*$, the position $\bm{p}^*$ can be predicted as:
 \begin{equation}
 \hat{\bm{p}} = f(\bm{\tau}^*).
 \end{equation}
The uncertainty of the position estimation can be determined by different ways. For instance, as shown in Fig.~\ref{fig:fig1}, we get different predictions from $k$ different trees, which are denoted by $\hat{\bm{p}}_1, \ldots,\hat{\bm{p}}_k$. Then the uncertainty can be determined as the variance of the $k$ predictions, yielding
\begin{equation}
	\bm{v}(\hat{\bm{p}} ) = \frac{1}{(k-1)} \sum_{i=1}^k (\hat{\bm{p}}_i-\hat{\bm{p}} )^2. 
\end{equation}
The above mentioned method to evaluate the uncertainty is based on the multiple decision trees in the RF algorithm. In ML, there are some other different ways of evaluating the reliability/uncertainty of the prediction. For example, the CNK method which measures the differences between the position predictions and the ones from a K-Nearest Neighbor (KNN) learner: 
\begin{equation}
	\bm{v}_{CNK}(\hat{\bm{p}} ) = \begin{bmatrix}(\hat{p}_{\textit{x}}-\hat{p}_{\textit{x}}^{knn})^2\\
	(\hat{p}_{\textit{y}}-\hat{p}_{\textit{y}}^{knn})^2
	\end{bmatrix},
\end{equation}
where $\hat{\bm{p}}^{knn}\triangleq[\hat{p}_{\textit{x}}^{knn},\hat{p}_{\textit{y}}^{knn}]^T$ denotes the 2-Dimensional position estimation from the KNN learner. In what follows, we will present the performance comparison between different methods. 
\begin{figure}[t]
	\centering
	\includegraphics[width=8.5cm]{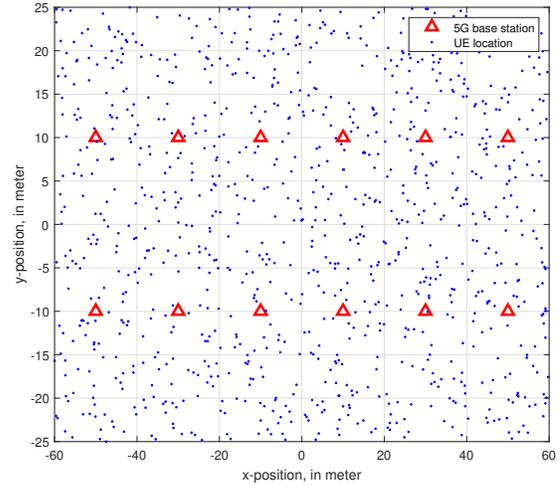}
	\caption{Indoor open office deployment.}
	\label{fig:fig2}
\end{figure} 

\begin{figure}[tb]
	\centering
	\includegraphics[width=8.5cm]{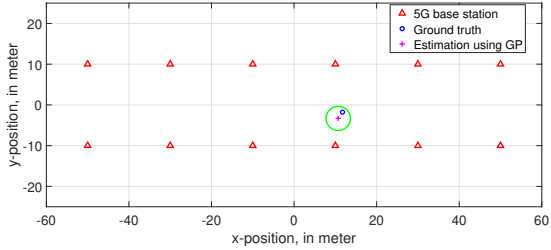}
	\caption{Uncertainty estimation (green circle) for one specific UE location.}
	\label{fig:fig4}
\end{figure}

\begin{figure}[tb]
	\centering
	\includegraphics[width=8.5cm]{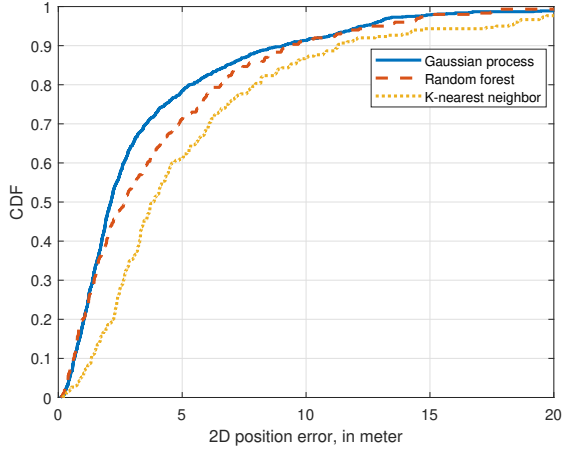}
	\caption{CDF of RMSE.}
	\label{fig:fig3}
\end{figure} 

\begin{figure*}[ht!]
     \centering
     \begin{subfigure}[]
         \centering
         \includegraphics[width=5cm]{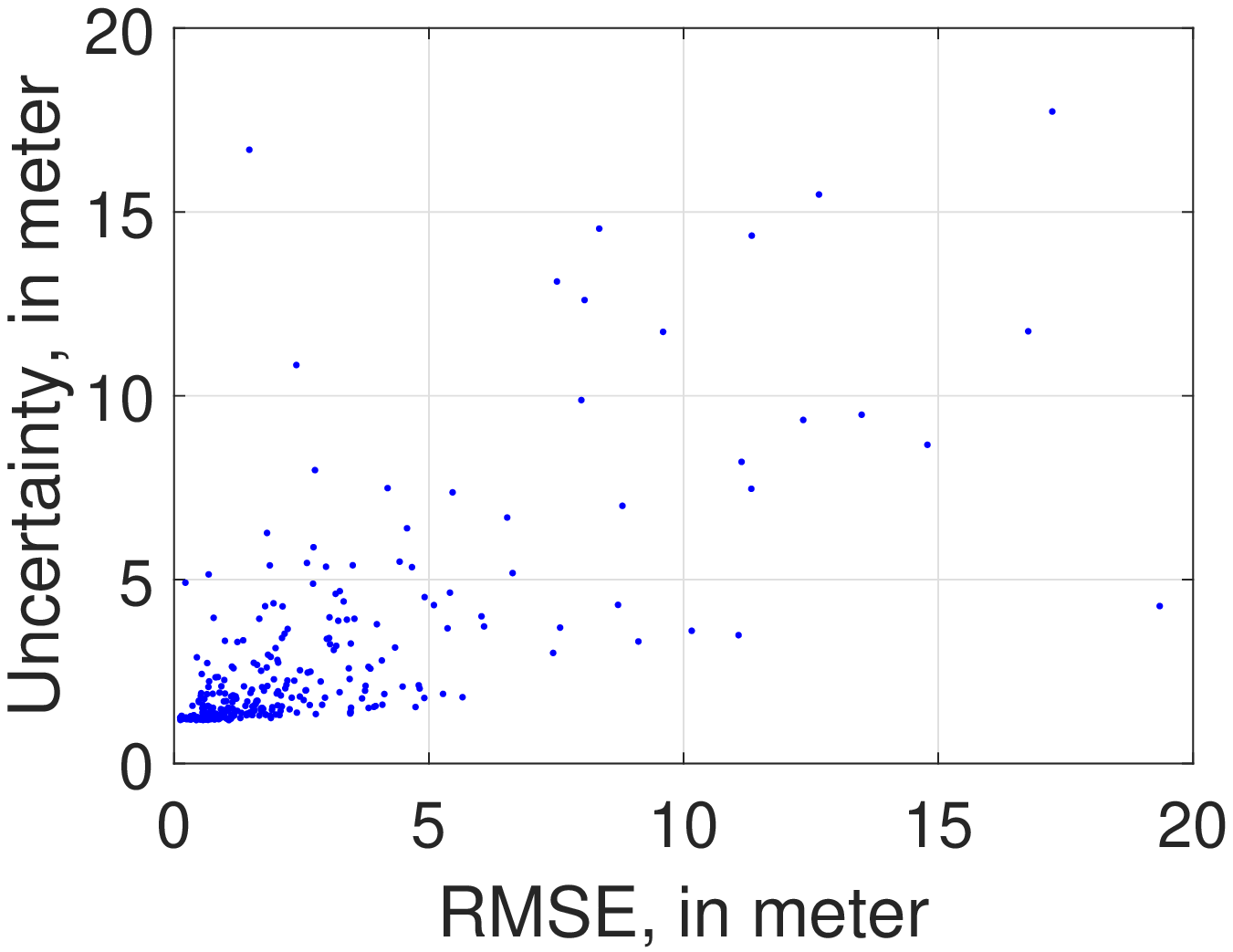}
         \label{fig:figure5a}
     \end{subfigure}
     \hfill
     \begin{subfigure}[]
         \centering
         \includegraphics[width=5cm]{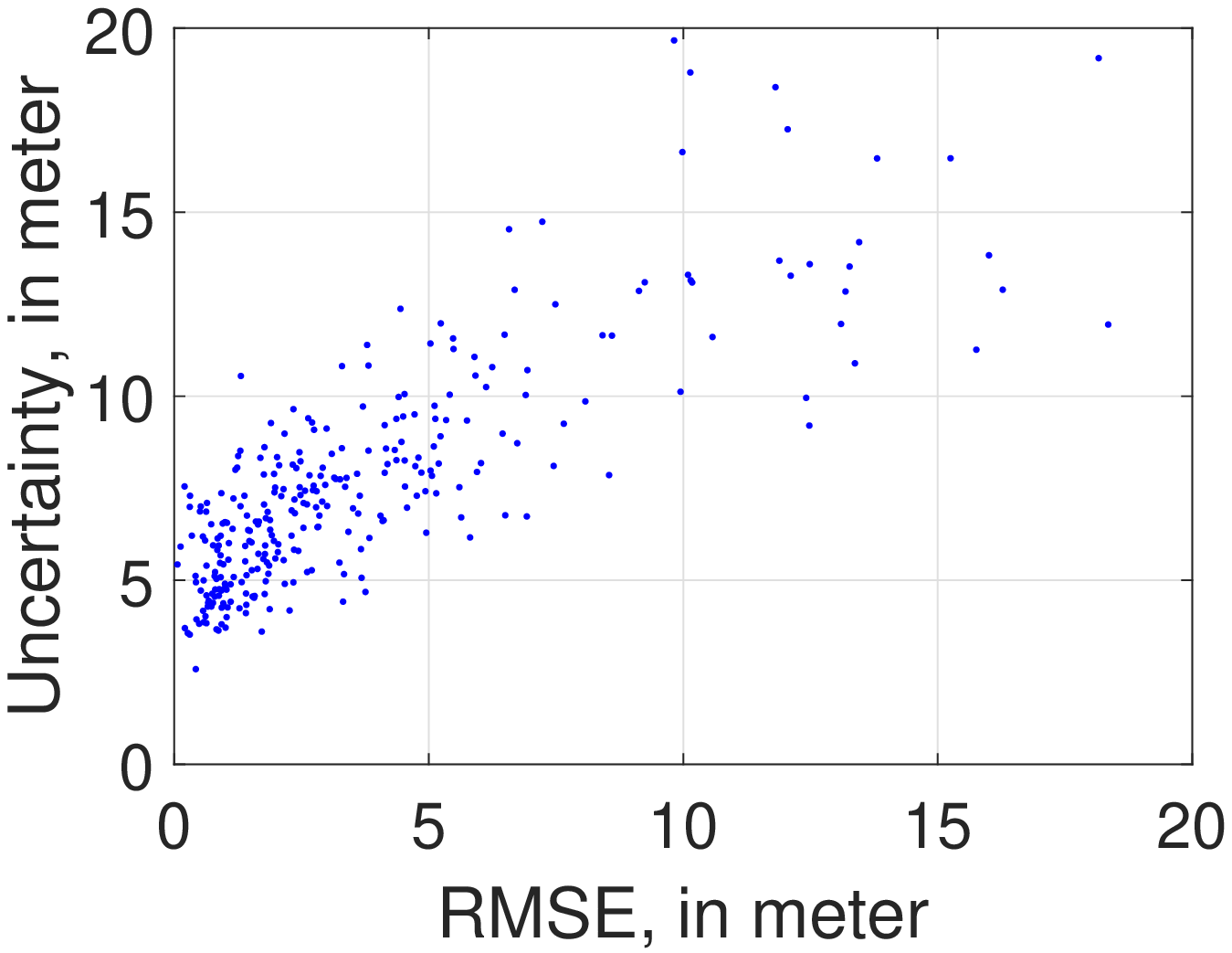}
         \label{fig:figure5b}
     \end{subfigure}
     \hfill
     \begin{subfigure}[]
         \centering
         \includegraphics[width=5cm]{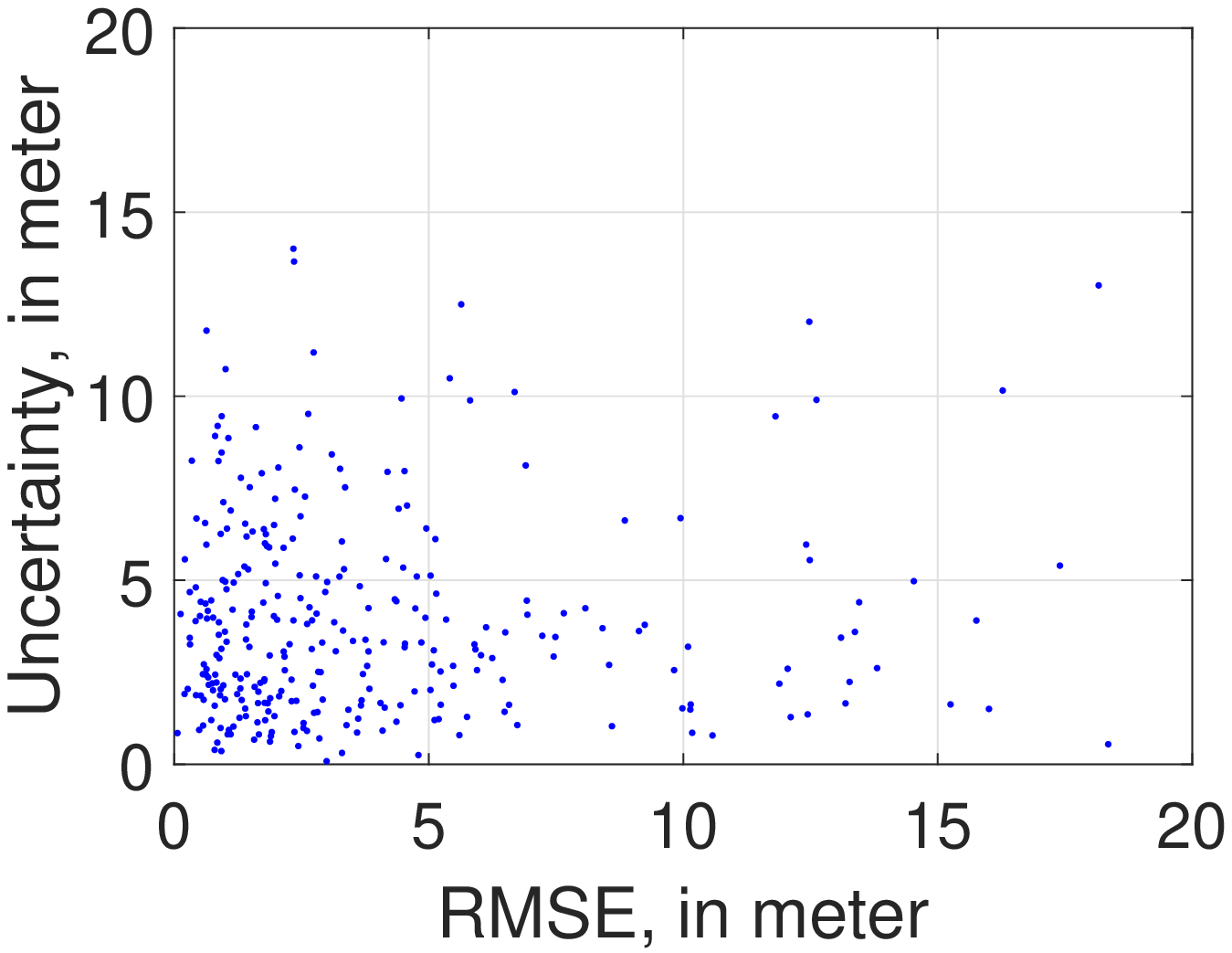}
         \label{fig:figure5c}
     \end{subfigure}
        \caption{RMSE versus uncertainty: (a) Gaussian Process, (b) Random Forest, (c) Random Forest CNK}
        \label{fig:figure5}
\end{figure*}

\section{Results}\label{sec:results}
In this section, the detailed description of the data set are provided first, and the positioning results and uncertainty assessment of the proposed methods are given.  
\subsection{Data Description}
In this work, we consider an indoor open office scenario with a total number of 12 5G BSs deployed in the area. The BSs are mounted on the ceiling at a height of 3m. The size of the area and the locations of BSs are illustrated in Fig.~\ref{fig:fig2} \cite{ioo_deployment}. In total, 1000 UEs (at height 1.5m) are randomly dropped within the deployment area. In the simulation, UEs measure the DL PRS signals and the ToA measurements between each UE and all 12 BSs are estimated and recorded. The estimated ToA corresponds to the time of flight of the PRS transmitted from a BS and reaching to UE via the first significant path.

\subsection{Simulation Results}
After the simulated data set has been generated, the two proposed ML approaches are applied to get both position estimates and the uncertainties associated with the estimations. To get a KNN prediction, the number of neighbors is selected as $3$. For the GP approach, the kernel function used here is Squared Exponential (SE) kernel:
\begin{equation}
    k(\tau, \tau') = \sigma_k^2\exp(-\frac{||\tau-\tau'||^2}{2l^2}),
\end{equation}
where $\sigma_k$ and $l$ are hyperparameters, which can be estimated by maximizing the likelihood function in \eqref{eq:likelihood}. For RF approach, the hyperparameters such as the number of trees are tuned through cross validation. 

The uncertainty assessments introduced in previous sections are given in $x$ and $y$ coordinates respectively. To have an overall representation of the uncertainty estimation performance, we propose using the following metric:
\begin{equation}
    c(\hat{\bm{p}})=\sqrt{v_x+v_y},
\end{equation}
where $\bm{v}(\hat{\bm{p}})\triangleq[v_x,v_y]^T$, $v_x$ and $v_y$ denote the uncertainty in $x$ and $y$ coordinate, respectively. Then, the uncertainty performance is measured by computing the correlation between the uncertainty value and the Root-Mean-Square Error (RMSE) of the position estimation. In Fig. \ref{fig:fig4}, the uncertainty estimation for one specific UE location is illustrated. Both the ground truth position and the estimated UE position using GP method are shown. The green circle indicates the uncertainty area around the estimated UE location. It is worth noting that the uncertainty area very well captures the true location of the UE.   

In Fig. \ref{fig:fig3}, we compare the Cumulative Distribution Function (CDF) of the RMSE for different ML approaches. The performance of KNN is also provided in the figure, since the position estimation from the KNN learner will be used later to compute the uncertainty using CNK method. GP has the best positioning performance over RF and K-nearest neighbor. The uncertainty estimation are plotted in Fig. \ref{fig:figure5} versus the RMSE. It can be seen that both GP and RF provide better correlation between the uncertainty and RMSE. To quantify the correlation/similarity between the uncertainty and RMSE, we compute and summarize the correlation between the uncertainty assessment and the true RMSE in table \ref{tab:table1}. The correlation coefficients varies between $0$ and $1$, with $1$ indicates the strongest correlation. As the correlation coefficient gets closer to $1$, it indicates more accurate uncertainty estimation.  

From above results, RF has the best uncertainty estimation performance. GP provides better positioning accuracy, but not as good as RF in uncertainty estimation. There is a trade-off between the positioning accuracy and the uncertainty estimation using GP and RF. The CNK method provides the least correlated uncertainty. This is expected from the CDF curve of the positioning accuracy. Since in the CNK method, the uncertainty is measured as the distance to the KNN prediction. To have better uncertainty estimation, the KNN prediction should be close to the ground truth position, which is not the case in this study.

\begin{table}[tb]
	\scriptsize
	\centering
	\caption{Uncertainty performance.} 
	\begin{tabular}[htdp]{lll}
		\hline
		Method                 & Correlation coefficient\\
		\hline
		Gaussian Process           & $0.6696$  \\
		Random Forest                     & $0.8465$   \\
		Random Forest CNK                     & $0.4911$\\
		\hline
	\end{tabular}
	\label{tab:table1}
\end{table}
\section{Conclusions}
\label{sec:conclusions}
In this work, we have proposed ML methods for uncertainty assessment in positioning. Two ML methods are introduced to perform position estimation and uncertainty assessment at the same time. The methods are evaluated using simulated ToA measurements from DL PRS transmissions. Results have shown that both GP and RF achieve satisfactory positioning accuracy, and the uncertainty assessment gives good indication of the reliability of the position prediction. In the future work, we are interested in considering the uncertainty in ToA measurements and how they may impact the uncertainty in position estimation. In addition, the impact of ground truth accuracy in the training data set is to be investigated. 

\section*{acknowledgments}
This work has been supported in parts by Ericsson Research AB and by the European Union’s Horizon 2020 research and innovation programme under grant agreement No.871249 (Research and Innovation Action), LOCalization and analytics on-demand embedded in the 5G ecosystem for Ubiquitous vertical applicationS (LOCUS).

\bibliographystyle{IEEEtran}
\bibliography{ref}

\begin{thebibliography}{10}
\providecommand{\url}[1]{#1}
\csname url@samestyle\endcsname
\providecommand{\newblock}{\relax}
\providecommand{\bibinfo}[2]{#2}
\providecommand{\BIBentrySTDinterwordspacing}{\spaceskip=0pt\relax}
\providecommand{\BIBentryALTinterwordstretchfactor}{4}
\providecommand{\BIBentryALTinterwordspacing}{\spaceskip=\fontdimen2\font plus
\BIBentryALTinterwordstretchfactor\fontdimen3\font minus
  \fontdimen4\font\relax}
\providecommand{\BIBforeignlanguage}[2]{{%
\expandafter\ifx\csname l@#1\endcsname\relax
\typeout{** WARNING: IEEEtran.bst: No hyphenation pattern has been}%
\typeout{** loaded for the language `#1'. Using the pattern for}%
\typeout{** the default language instead.}%
\else
\language=\csname l@#1\endcsname
\fi
#2}}
\providecommand{\BIBdecl}{\relax}
\BIBdecl

\bibitem{8667173}
Q.~K. Ud~Din~Arshad, A.~U. Kashif, and I.~M. Quershi, ``A review on the
  evolution of cellular technologies,'' in \emph{2019 16th International
  Bhurban Conference on Applied Sciences and Technology (IBCAST)}, 2019, pp.
  989--993.

\bibitem{7970319}
W.~Lin and R.~W. Ziolkowski, ``Compact, omni-directional, circularly-polarized
  mm-wave antenna for device-to-device (d2d) communications in future 5g
  cellular systems,'' in \emph{2017 10th Global Symposium on Millimeter-Waves},
  2017, pp. 115--116.

\bibitem{9217500}
S.~Gyawali, S.~Xu, Y.~Qian, and R.~Q. Hu, ``Challenges and solutions for
  cellular based v2x communications,'' \emph{IEEE Communications Surveys
  Tutorials}, vol.~23, no.~1, pp. 222--255, 2021.

\bibitem{4127518}
T.~Kos, M.~Grgic, and G.~Sisul, ``Mobile user positioning in gsm/umts cellular
  networks,'' in \emph{Proceedings ELMAR 2006}, 2006, pp. 185--188.

\bibitem{8377447}
S.~M. Razavi, F.~Gunnarsson, H.~Ryden, A.~Busin, X.~Lin, X.~Zhang, S.~Dwivedi,
  I.~Siomina, and R.~Shreevastav, ``Positioning in cellular networks: Past,
  present, future,'' in \emph{2018 IEEE Wireless Communications and Networking
  Conference (WCNC)}, 2018, pp. 1--6.

\bibitem{9053157}
J.~A. del Peral-Rosado, F.~Gunnarsson, S.~Dwivedi, S.~M. Razavi, O.~Renaudin,
  J.~A. López-Salcedo, and G.~Seco-Granados, ``Exploitation of 3d city maps
  for hybrid 5g rtt and gnss positioning simulations,'' in \emph{ICASSP 2020 -
  2020 IEEE International Conference on Acoustics, Speech and Signal Processing
  (ICASSP)}, 2020, pp. 9205--9209.

\bibitem{8964963}
Y.~Cheng and T.~Zhou, ``Uwb indoor positioning algorithm based on tdoa
  technology,'' in \emph{2019 10th International Conference on Information
  Technology in Medicine and Education (ITME)}, 2019, pp. 777--782.

\bibitem{8877160}
R.~Keating, M.~Säily, J.~Hulkkonen, and J.~Karjalainen, ``Overview of
  positioning in 5g new radio,'' in \emph{2019 16th International Symposium on
  Wireless Communication Systems (ISWCS)}, 2019, pp. 320--324.

\bibitem{7217158}
H.~Ryden, S.~M. Razavi, F.~Gunnarsson, S.~M. Kim, M.~Wang, Y.~Blankenship,
  A.~Grovlen, and A.~Busin, ``Baseline performance of lte positioning in 3gpp
  3d mimo indoor user scenarios,'' in \emph{2015 International Conference on
  Localization and GNSS (ICL-GNSS)}, 2015, pp. 1--6.

\bibitem{8970252}
G.~Destino, T.~Mahmoodi, R.~Shreevastav, D.~Shrestha, and I.~Siomina, ``A new
  position quality metric for nr rat dependent otdoa positioning methods,'' in
  \emph{2019 16th Workshop on Positioning, Navigation and Communications
  (WPNC)}, 2019, pp. 1--5.

\bibitem{2021arXiv210203361D}
S.~{Dwivedi}, R.~{Shreevastav}, F.~{Munier}, J.~{Nygren}, I.~{Siomina},
  Y.~{Lyazidi}, D.~{Shrestha}, G.~{Lindmark}, P.~{Ernstr{\"o}m}, E.~{Stare},
  S.~M. {Razavi}, S.~{Muruganathan}, G.~{Masini}, {\r{A}}.~{Busin}, and
  F.~{Gunnarsson}, ``{Positioning in 5G networks},'' \emph{arXiv e-prints}, p.
  arXiv:2102.03361, Feb. 2021.

\bibitem{8417725}
H.~Ryden, S.~M. Razavi, F.~Gunnarsson, and I.~Olofsson, ``Cellular network
  positioning performance improvements by richer device reporting,'' in
  \emph{2018 IEEE 87th Vehicular Technology Conference (VTC Spring)}, 2018, pp.
  1--5.

\bibitem{otdoa_sven}
S.~Fischer, ``{Observed Time Difference of Arrival (OTDOA) Positioning in 3GPP
  {LTE}},'' {Qualcomm Technologies Inc.)}, Tech. Rep., June 2014.

\bibitem{BOSNIC2008504}
Z.~Bosnić and I.~Kononenko, ``Comparison of approaches for estimating
  reliability of individual regression predictions,'' \emph{Data \& Knowledge
  Engineering}, vol.~67, no.~3, pp. 504--516, 2008.

\bibitem{RW06}
C.~E. Rasmussen and C.~K.~I. Williams, \emph{Gaussian Processes for Machine
  Learning}.\hskip 1em plus 0.5em minus 0.4em\relax Cambridge, MA, USA: MIT
  Press, 2006.

\bibitem{ioo_deployment}
3GPP, ``{Study on channel model for frequencies from 0.5 to 100 GHz (Release
  16)},'' {3rd Generation Partnership Project (3GPP)}, Technical Report (TR)
  38.901, V16.1.0, 2019.

\end{thebibliography}
\end{document}